\documentclass{article}

\usepackage{amssymb}
\usepackage{graphicx}
\usepackage{dcolumn}
\usepackage{bm}
\date{}
\begin{document}

\title{Polymer quantization versus the Snyder noncommutative
space}
\author{M. A. Gorji$^{1}$\thanks{email:m.gorji@stu.umz.ac.ir},
\hspace{.2 cm} K. Nozari$^1$\thanks{{email: knozari@umz.ac.ir}}
\hspace{.2cm} and \hspace{.2cm}B. Vakili$^2$\thanks{email:
b.vakili@iauctb.ac.ir}\vspace{.2cm}\\
$^1${\small {\it Department of Physics, Faculty of Basic Sciences,
University of Mazandaran},}
\\{\small {\it P.O. Box 47416-95447, Babolsar, Iran.}}\\$^2${\small
{\it Department of Physics, Central Tehran Branch, Islamic Azad
University, Tehran, Iran}}}

\maketitle
\begin{abstract}
We study a noncanonical Hilbert space representation of the polymer
quantum mechanics. It is shown that Heisenberg algebra get some
modifications in the constructed setup from which a generalized
uncertainty principle will naturally come out. Although the
extracted physical results are the same as those obtained from the
standard canonical representation, the noncanonical representation
may be notable in view of its possible connection with the
generalized uncertainty theories suggested by string theory. In this
regard, by considering an Snyder-deformed Heisenberg algebra we show
that since the translation group is not deformed it can be
identified with a polymer-modified Heisenberg algebra. In classical
level, it is shown the noncanonical Poisson brackets are related to
their canonical counterparts by means of a Darboux transformation on
the corresponding phase space.
\vspace{5mm}\\
\begin{description}
\item[PACS numbers]
04.60.Nc, 04.60.Ds, 04.60.Pp
\item[Key Words]
Polymer quantization, Noncommutative geometry, Snyder space
\end{description}
\end{abstract}

\section{Introduction}

When one takes into account the gravity and quantum theory issues in
the special relativity, the general theory of relativity and quantum
field theory can be achieved respectively. However, the need for a
quantum theory of gravity is inevitable either in general relativity
to resolve the space-time singularities or in quantum field theories
to overcome the ultraviolet divergencies. Evidently, both of these
problems have the same origin and may be resolved by taking an
effective cutoff on ultraviolet, i.e., a minimal observable length
scale (for singularity resolution in general relativity see
\cite{Ashtekar} and for the ultraviolet regularization of the
quantum field theory see \cite{Kempf}). This issue is also proposed
by the quantum gravity candidates such as loop quantum gravity
\cite{LQG} and string theory \cite{ST} which both of them suggest
some modifications to the structure of the standard quantum
mechanics. From string theory point of view, a minimum length scale
$\alpha$ which is responsible for the strings' size \cite{GUP}, will
appear due to the corrections to the standard uncertainty principle.
Such generalized uncertainty relations can also be realized from a
noncanonical (deformed) Heisenberg algebra on the corresponding
Hilbert space \cite{GUP-HS,GUP-HS2}. Noncommutativity between
space-time coordinates was first introduced by Snyder in 1947
\cite{Snyder}, which was a Lorentz invariant theory with a minimal
length scale in its formalism. The associated noncanonical
Heisenberg algebra is also very similar to the one obtained in the
context of the string theory \cite{Mignemi}.

In recent years, another formalism of quantum mechanics, called
polymer quantum mechanics, is proposed which supports the idea of
the existence of a minimal length scale $\mu$, known as the polymer
length, without any attribution to the noncanonical Heisenberg
algebra \cite{QPR}. In this approach for quantization of a given
dynamical system one uses methods very similar to the effective
models of loop quantum gravity. Here, the main role is played by the
mentioned polymer length scale in such a way that, unlike the
deformed algebraic structure usually coming from the noncommutative
phase-space variables, it enters into the Hamiltonian of the system
to deform its functional form into a new one called the polymeric
Hamiltonian. However, a question may be raised: can a
polymer-deformed quantum mechanical theory be compatible with a
theory with noncanonical (deformed) Heisenberg algebra in its formalism? In
this paper we are going to answer this question by concentrating on
the noncanonical Hilbert space representation of the polymer quantum
mechanics. In particular, we show that the noncanonical
representation of the polymer quantum mechanics coincides with an
Snyder-deformed (noncanonical) Heisenberg algebra just by
reasonable identification of the deformation parameters as
$\mu=\alpha\approx\,l_{_{\rm Pl}}$, where $l_{_{\rm
Pl}}\sim\,10^{-33}\,cm$ is the Planck length.

The structure of the paper is as follows: In section 2, general
formulation of the polymer quantization is briefly reviewed. The
standard canonical representation of the polymer quantum mechanics
is presented in section 3. In section 4, the noncanonical
representation of the polymer quantum mechanics is investigated and
the corresponding Hilbert space representation is also studied. The
generalized uncertainty principle is obtained in this setup and the
simple harmonic oscillator is considered as a relevant example. In
section 5, inspired by the theory of Snyder noncommutative spaces,
we first construct a noncanonical Heisenberg algebra and then show
that it can be identified with the polymer-modified Heisenberg
algebra obtained from the noncanonical polymeric representation. The
classical limits of the canonical and noncanonical polymeric
representation is the subject of section 6 in which it is shown that
these two setups are related to each other just by a Darboux
transformation on the associated phase space. Section 7 is devoted
to the summary and conclusions.

\section{Polymer Quantization}

Polymer quantum mechanics is an alternative framework of the
commutation relation that is investigated in the context of the loop
quantum gravity \cite{QPR}. In contrast to the standard
Schr\"{o}dinger picture, this representation gives an ultraviolet
cutoff due to the existence of the polymer length as a possible
minimum length scale for the system under consideration. In this
section, we briefly review the kinematics and dynamics of the
polymer quantum mechanics (see Refs. \cite{QPR,QPR22,QPR2,Campiglia}
for more details).

\subsection{Kinematics}

Consider a classical mechanical system consisting of a single
particle which is characterized by the position ${\tilde q}$ and its
conjugate momentum ${\tilde p}$ (we will use the notation $({\tilde
q},{\tilde p})$ for the classical variables through this paper). In
the two-dimensional phase space $({\tilde q},{\tilde p})$, these
quantities satisfy the ordinary canonical Poisson algebra
\begin{equation}\label{CPA}
\{\tilde{q},\tilde{p}\}=1\,.
\end{equation}
Canonical quantization is straightforward. Classical variables
$({\tilde q},{\tilde p})$ are replaced by quantum operators $(q,p)$
and the role of the Poisson brackets will be played by the Dirac
commutators. Then one is led to the quantum Heisenberg algebra
\begin{equation}\label{CCR}
[q,p]=i\,,
\end{equation}
in which we have set $\hbar=1$. The operators $(q,p)$ are unbounded
and the associated Hilbert space should necessarily be
infinite-dimensional. However, in quantum gravity regime in order to
take the natural cutoffs such as minimal length and maximal momentum
into account, one deals with bounded variables. One way to out of
this difficulty is implementing the Weyl algebra \cite{QPR,QPR2}.
Thus, in an alternative way, but more admissible for the polymer
representation, one can adopt the complex exponential versions of
the operators $q$ and $p$ (Weyl operators) as
\begin{eqnarray}\label{Exp-Op}
U(\alpha)=e^{i\alpha q},\hspace{1cm}V(\beta)=e^{i\beta p}\,,
\end{eqnarray}
with
\begin{eqnarray}\label{Weyl-AB}
U(\alpha_1).U(\alpha_2)=U(\alpha_1+\alpha_2),\hspace{1cm}
V(\beta_1).V(\beta_2)=V(\beta_1+\beta_2)\,.
\end{eqnarray}
The parameters $\alpha$ and $\beta$ have dimensions of momentum and
length respectively. Using the Baker-Hausdorff lemma, the Heisenberg
algebra (\ref{CCR}) for the operators $(q,p)$ takes the following
Weyl algebra form
\begin{eqnarray}\label{Weyl-A}
U(\alpha).V(\beta)=e^{-i\alpha\beta}\,V(\beta).U(\alpha)\,,
\end{eqnarray}
generated by the exponentiated position and momentum operators
(\ref{Exp-Op}). Then, the ordinary Schr\"{o}dinger representation
can be obtained via the standard Gel'fand-Naimark-Segal construction
\cite{QPR}. In position polarization of the Schr\"{o}dinger
representation, the corresponding Hilbert space is
${\mathcal H}_{\rm Sch}=L^2({\mathbb R},dq)$ which is the space of
the square integrable functions with respect to the Lebesgue measure
$dq$ on the real line ${\mathbb R}$. Therefore, the Hilbert space in
the momentum polarization will be
\begin{equation}\label{HS}
{\mathcal H}_{_{\rm Sch}}=L^2({\mathbb R},dp)\,,
\end{equation}
in which the operators are represented as
\begin{eqnarray}\label{MHS}
p.\psi(p)=p\psi(p),\hspace{1cm}q.\psi(p)=i\frac{\partial\psi(p)}{\partial p}\,,
\end{eqnarray}
where $\psi(p)=\langle p|\psi\rangle$ is a typical wave function.
According to the Stone-Von Neumann uniqueness theorem every
irreducible representation of the Weyl algebra (\ref{Weyl-A}) which
is weakly continuous in the parameters $\alpha$ and $\beta$, is
unitarily equivalent to the standard Schr\"{o}dinger representation.
But, the so-called polymer representation is unitarily inequivalent
to the Schr\"{o}dinger one \cite{QPR}. Evidently, the construction
should violates at least one assumption of the Stone-Von Neumann
uniqueness theorem. Indeed, the exponentiated momentum operator
$V(\beta)$ is no longer weakly continuous in $\beta$ in the polymer
representation. Taking this hypothesis into account, the polymer
representation can be obtained from the the Weyl algebra
(\ref{Weyl-A}) via the Gel'fand-Naimark-Segal construction
\cite{QPR,QPR2}.

In this regard, we define an abstract ket $|\lambda\rangle$, labeled
by a real number $\lambda$, which shall belong to the non-separable
Hilbert space ${\mathcal{H}}_{\rm poly}$. An appropriate state can be
obtained by taking a linear combination of a finite collections of
vectors $|\lambda_i\rangle$ where $\lambda_i\in\,{\mathbb R}$ and
$i=1,2, ...,N$ as
\begin{equation}\label{state}
|\psi\rangle=\,\sum_{i=1}^{N}a_i\,|\lambda_i\rangle\,.
\end{equation}
Then, the polymer inner product between the fundamental kets will be
\begin{equation}\label{P-IP}
\langle\lambda|\nu\rangle=\delta_{\lambda,\nu}\,.
\end{equation}
One can work in position or momentum polarization. In momentum
polarization, the states are denoted by $\psi(p)=\langle\,p|\psi\rangle$
where
\begin{eqnarray}\label{pstate}
\psi_{\lambda}(p)=\langle\,p|\lambda\rangle=e^{i\lambda p}.
\end{eqnarray}
Since here the operator $V(\beta)$, is not weakly continuous in
$\beta$, the associated momentum operator $p$ does not exist.
However, $V(\beta)$ itself is a well-defined operator which acts on
the states (\ref{pstate}) as
\begin{equation}\label{exp-p}
V(\nu).\,\psi_{\lambda}(p)=e^{i\nu p}\,e^{i\lambda p}=
e^{i(\lambda+\nu)p}=\psi_{\lambda+\nu}(p)\,.
\end{equation}
On the other hand the operator $q$ is well-defined in the sense that
the corresponding exponential operator
$U(\alpha)$ is continuous in $\alpha$. Therefore, the operator $q$
can be identified as
\begin{eqnarray}\label{exp-q}
q.\psi_{\lambda}(p)=i\frac{\partial}{\partial p}\psi_{\lambda}(p)
=-\lambda e^{i\lambda p}=-\lambda\psi_{\lambda}(p)\,,
\end{eqnarray}
which can be interpreted as a discrete operator in the sense that
its eigenvalues are labeled by $\lambda$s which in turn, may take
their values from a continuum such that the states (\ref{pstate})
are orthonormal. Now, what remains is the definition of an inner
product with respect to an appropriate measure on the abstract
Hilbert space ${\mathcal H}_{\rm poly}$. The above definitions
correspond to the Bohr compactification of the real line ${\mathbb
R}_{_B}$ which is a compact group, and the Haar measure $d\mu_{_H}$
is a natural measure defined on it. Therefore, the polymeric Hilbert
space ${\mathcal H}_{\rm poly}$ in momentum polarization will be
\cite{QPR2}
\begin{equation}\label{PHS}
{\mathcal H}_{\rm poly}=L^2({\mathbb R}_{_B},d\mu_{_H})\,.
\end{equation}
Therefore, one may define the following inner product for the
periodic functions $\psi_{\lambda}(p)$ on this Hilbert space,
\begin{eqnarray}\label{PHS-IP}
\langle\psi_{\lambda'}|\psi_{\lambda}\rangle_{\rm poly}=\int_{{\mathbb R}_{_B}}
d\mu_{_H}\psi^{\ast}_{\lambda'}(p)\psi_{\lambda}(p):=\lim_{L
\rightarrow\infty}\frac{1}{2L}\int_{-L}^{+L}dp\,\psi^{\ast}_{\lambda'}(p)
\psi_{\lambda}(p)=\delta_{\lambda,\lambda'}\,.
\end{eqnarray}
In contrast to the standard Schr\"{o}dinger representation,
the Dirac delta function is replaced by the Kronecker delta
in this setup.

\subsection{Dynamics}
Our starting point in this section is to consider a classical
Hamiltonian function and see how one can passes from this point to
the polymeric Hamiltonian. To do this, let us begin with the
classical nonrelativistic Hamiltonian function
\begin{equation}\label{nhamiltonian}
H=\frac{{\tilde p}^2}{2m}+W({\tilde q})\,,
\end{equation}
where $m$ is a mass parameter and $W$ denotes a potential function.
At classical level, the pair $({\tilde q}, {\tilde p})$ satisfy the
Poisson algebra (\ref{CPA}) and at quantum level, they should be
replaced with their operator counterpart $(q,p)$ and the standard
Heisenberg algebra (\ref{CCR}). Both of the position $q$ and
momentum $p$ are well-defined operators in the standard
Schr\"{o}dinger representation since their exponentiated operators
$U(\alpha)$ and $V(\beta)$ are weakly continuous in $\alpha$ and
$\beta$ respectively. In momentum polarization, the position and
momentum operators are given by the relations (\ref{MHS}) on the
Hilbert space (\ref{HS}). In polymer representation, however, this
argument fails to be applicable since the exponentiated momentum
operator $V(\beta)$ is no longer weakly continuous in $\beta$ and
consequently the corresponding momentum operator does not exist. The
main task now is to define a suitable momentum operator and its
square in polymer framework to achieve a quantum Hamiltonian
operator from the classical version (\ref{nhamiltonian}). The
situation is very similar to that in loop quantum gravity, where
while the connection is not a well-defined operator, its holonomy is
and plays the role of the con- jugate momentum for the volume
operator with discrete eigenvalues \cite{Ashtekar2}. The standard
prescription is to define the lattice $\gamma_{\mu}$ on the
configuration space as
\begin{equation}\label{lattice}
\gamma_\mu=\{q\in{\mathbb{R}}|\,q=n\mu,\,\forall\,n\in{\mathbb Z}\},
\end{equation}
which ensures the self-adjointness of the position operator and
also discreteness of its eigenvalues. In order to regulate a
momentum operator in this setup, we consider the action of its
exponentiated operator $V(\beta)=e^{i\beta p}$ on the basic kets
as
\begin{eqnarray}\label{PLSO}
V(\mu)\,|\mu_n\rangle=e^{i\mu p}e^{i\mu_n p}=e^{i(\mu+\mu_n)p}
=\,|\mu_n+\mu\rangle=|\mu_{n+1}\rangle.
\end{eqnarray}
The physical states in the separable Hilbert space
${\mathcal{H}}_{\gamma_\mu}$ are then of the form
\begin{equation}\label{PLS}
|\psi\rangle=\,\sum_n b_n |\mu_n\rangle\,,
\end{equation}
with coefficients $b_n$ satisfy $\sum_n|b_n|^2<\infty$. In this
respect, one can regulate the momentum operator in polymer framework
by means of the exponential shift operator as \cite{QPR}
\begin{eqnarray}\label{PLMO}
p_\mu\,|\mu_n\rangle=\,\frac{1}{2i{\mu}}\Big(V(\mu)
-V(-\mu)\Big)|\mu_n\rangle=\frac{1}{2i\mu}
\Big(|\mu_{n+1}\rangle-|\mu_{n-1}\rangle\Big)\,.
\end{eqnarray}
The squared momentum operator can be defined in the same way just by
twice acting the polymeric momentum operator (\ref{PLMO}) on the
states as \footnote{The operator $p_\mu.p_\mu$ shifts the states two
steps in the lattice to both sides. However, one can rescale this
operator as $\mu\rightarrow\,\mu/2$ to give an approximation for the
squared momentum operator that shifts the states one step in lattice
\cite{QPR,QPR2}.}
\begin{eqnarray}\label{PLMO2}
p_{\mu}^2\,|\mu_n\rangle=\,\frac{1}{4\mu^2}\Big(2-V(2\mu)
-V(-2\mu)\Big)|\mu_n\rangle=\frac{1}{4\mu^2}
\Big(2|\mu_n\rangle-|\mu_{n+2}\rangle-|\mu_{n-2}\rangle
\Big).
\end{eqnarray}
Then, the quantum polymeric Hamiltonian operator can be deduced
by means of the polymeric squared momentum operator (\ref{PLMO2})
from the classical Hamiltonian function (\ref{nhamiltonian}) as
\begin{equation}\label{QH}
H_{\mu}=\frac{1}{8m\mu^2}\Big(2-V(2\mu)-V(-2\mu)\Big)+W(q)\,.
\end{equation}
Evidently, the momentum is periodic in the sense that the functions
of $p$ that preserve the lattice $\gamma_\mu$ are of the form
$e^{im\mu p}$ for integer $m$. There is a maximal value for the
momentum which is defined by the lattice spacing $\mu$. Thus the
momentum operator is bounded self-adjoint operator on the lattice.
In classical regime, it means that the momentum part of the phase
space is compactified to a circle $S^1$ (see the section for the
classical phase space). Thus, the separable Hilbert space
${\mathcal H}_{\gamma_\mu}$ in momentum polarization will be
\cite{QPR2}
\begin{equation}\label{HS-S1M}
{\mathcal H}_{\gamma_\mu}=L^2(S^1,d\mu_{S^1})\,,
\end{equation}
where $d\mu_{S^1}$ is now the Haar measure for $S^1$. This Hilbert
space is a subspace of the nonseparable polymeric Hilbert space
(\ref{PHS}). One can also define the momentum operator and its
square on the nonseparable Hilbert space ${\mathcal H}_{\rm Poly}$,
but their physical interpretation is problematic \cite{QPR2}. For
instance, as we will see, the position eigenvectors are orthonormal
only when one considers the lattice (\ref{lattice}) on the separable
Hilbert space (\ref{HS-S1M}).

Like the standard Schr\"{o}dinger representation, one can solve the
Hamiltonian eigenvalue problem in position or momentum polarization
in polymer setup. But, we prefer to work in momentum polarization in
this paper. Indeed, the Hamiltonian eigenvalue problem
\begin{equation}\label{HEVP}
H_{\mu}|\psi\rangle=E|\psi\rangle\,,
\end{equation}
takes the form of a difference equation for the Fourier coefficient of a general state $\psi(q)$
in position polarization and it becomes a differential equation when it acts on the state
(\ref{PLS}) in momentum polarization of the Hilbert space ${\mathcal{H}}_{\gamma_\mu}$ (see
Refs. \cite{QPR,QPR3,Barbero} for more details).

\section{Canonical Polymeric Representation}

Using definition (\ref{Exp-Op}) for the shift operator
$V(\mu)=\,e^{i\mu p}$ in the relation (\ref{PLMO}), gives an
approximation for the polymeric momentum as
\begin{eqnarray}\label{Polymerization}
p_\mu\approx\frac{1}{\mu}\sin(\mu\,p),
\end{eqnarray}
with $p\in(-\pi/\mu,+\pi/\mu)$. Note also that this relation implies
a bounded range for the polymeric momentum as
$p_\mu\in(-1/\mu,+1/\mu)$. In the limit of $\mu p \ll1$, the
polymeric momentum operator $p_{\mu}$ coincides with its standard
unbounded canonical one $p\in(-\infty,+\infty)$. This result shows
that the compactification of the momentum by the circle $S^1$
induces an ultraviolet cutoff defined by the lattice spacing. In the
same way, the polymeric Hamiltonian operator (\ref{QH}) can be
approximated as
\begin{equation}\label{qhamiltonian}
H_\mu=\,\frac{1}{4m\mu^2}\Big(1\,-\,\cos(2\mu p)\Big)+W(q).
\end{equation}
Interestingly, this Hamiltonian gives both the maximal momentum and
also maximal energy for the case of a free particle \cite{QPR2}.

Note that one can represent the separable polymeric Hilbert space
(\ref{HS-S1M}) with either of the pairs $(q,p)$ and $(q,p_\mu)$ on
the lattice (\ref{lattice}). The pair $(q,p)$ satisfy the canonical
Heisenberg algebra (\ref{CCR}) and we call its components as
canonical operators. In the next section, we will show that the pair
$(q,p_\mu)$ satisfy a noncanonical Heisenberg algebra and
consequently we call them as noncanonical operators.

In the momentum polarization of the canonical polymeric representation of the quantum
mechanics, the position and momentum act as a derivative and multiplication operators
respectively
\begin{eqnarray}\label{PLMR-QP}
q.\psi(p)=i\frac{\partial}{\partial p}\psi(p)\,,\hspace{1cm}
p_\mu.\psi(p)=\frac{1}{\mu}\sin(\mu\,p)\psi(p).
\end{eqnarray}
Also, corresponding Hilbert space is nothing but (\ref{HS-S1M}) with
uniform measure
\begin{equation}\label{HS-S1}
{\mathcal H}_{\gamma_\mu,p}=L^2(S^1,dp)\,.
\end{equation}
Therefore, the inner product for the periodic functions
$\psi_\mu(p)$ on the Hilbert space (\ref{HS-S1}) can be obtained via
the definition (\ref{PHS-IP}) as
\begin{eqnarray}\label{IP-PLMR}
\langle\psi_1|\psi_2\rangle_{\gamma_\mu}=\frac{\mu}{2\pi}
\int_{-\pi/\mu}^{+\pi/\mu}\,dp\,\psi_1^{\ast}(p)\psi_2(p),
\end{eqnarray}
with $p\in(-\pi/\mu,\pi/\mu)$. The completeness relation then will
be
\begin{eqnarray}\label{CIO}
1=\frac{\mu}{2\pi}\int_{-\pi/\mu}^{+\pi/\mu}\,dp|p\rangle\langle p|\,,
\end{eqnarray}
and the inner product of momentum eigenvectors remains unchanged as
$\langle p|p'\rangle= \delta(p-p')$. The position eigenvalue problem
$q.\psi(p)=\lambda\psi(p)$ takes the form of a differential equation
$i\frac{\partial}{\partial p}\psi_\lambda(p)=\lambda\psi_\lambda(p)
$ with solution $\psi_\lambda(p)\sim\,e^{-i\lambda p}$. It is
important to note that the eigenvalues $\lambda$ here, should
necessarily be quantized with respect to the polymer length $\mu$
since the states (\ref{PLS}) leave the lattice (\ref{lattice})
invariant.

\section{Noncanonical Polymeric Representation}

As we have mentioned above, one may also work with the pair
$(q,p_\mu)$ on the separable Hilbert space (\ref{HS-S1M}). We note
that while the canonical momentum $p$ satisfies the canonical
Heisenberg algebra (\ref{CCR}), it is easy to show that the
polymeric momentum $p_\mu$ satisfies the noncanonical (deformed)
Heisenberg algebra
\begin{equation}\label{CR-PL}
[q,p_\mu]=i\sqrt{1-(\mu p_\mu)^2}\,.
\end{equation}
Because of this, we call the components of the pair $(q,p_\mu)$ as
the noncanonical operators. It is important to note that the relation
(\ref{CR-PL}) would be only considered on the lattice to ensure the
self-adjointness of the position and momentum operators. Nevertheless,
we consider this issue in more details to compare the results with
the case of the Snyder noncommutative space in the next section. In
the noncanonical representation the Hamiltonian operator takes the
standard functional form
\begin{equation}\label{EV-QH}
H_\mu=\frac{p_{\mu}^2}{2m}+W(q)\,.
\end{equation}
Thus, one may work within two equivalent pictures on the separable
Hilbert space (\ref{HS-S1M}): {\it i}) Working with the standard
canonical momentum $p$ which satisfies the standard Heisenberg
algebra (\ref{CCR}) together with the polymer-modified Hamiltonian
operator (\ref{qhamiltonian}). {\it ii}) Working with the polymeric
momentum $p_\mu$ which satisfies the polymer-modified Heisenberg
algebra (\ref{CR-PL}) and the corresponding Hamiltonian operator
(\ref{EV-QH}) with standard functional form. The former is the
standard canonical polymeric representation of the quantum mechanics
\cite{QPR,QPR2}. The latter, however, is the noncanonical polymeric
representation of quantum mechanics which we are going to consider
its Hilbert space representation in this paper.

\subsection{Hilbert Space Representation}

Clearly, this Hilbert space is unitarily equivalent to the Hilbert
space (\ref{HS-S1}). This is because they may be considered as the
different faces of the separable Hilbert space (\ref{HS-S1M}) and
consequently all the physical results should be the same.
Nevertheless, as we will show, the noncanonical representation
would be interesting due to its possible connection with the
theories based on the noncommutative Heisenberg algebra such as
the generalized uncertainty relations suggested by string theory
\cite{GUP,GUP-HS,GUP-HS2}. Also, the noncanonical representation
is more admissible from the statistical point of view when the
polymer quantization is applied to the thermodynamical systems
\cite{Gorji2,Gorji3}.

In momentum polarization, the position and the polymeric momentum
operators act on the physical states as
\begin{eqnarray}\label{Snyd-MHS}
p_\mu.\psi(p_\mu)=p_\mu\psi(p_\mu),\hspace{1cm}q.\psi(p_\mu)=i
\sqrt{1-(\mu p_\mu)^2}\frac{\partial\psi(p_\mu)}{\partial p_\mu},
\end{eqnarray}
where $\psi(p_\mu)=\langle p_\mu|\psi\rangle$ and $|\psi\rangle$ is
given by (\ref{PLS}). The requirement that the operators should be
self-adjoint in the noncanonical chart means that we have to define
an appropriate Haar measure on the Hilbert space (\ref{HS-S1M})
which will be
\begin{equation}\label{Snyd-HS}
{\mathcal H}_{_{\gamma_\mu,p_\mu}}=L^2\Big(S^1,dp_\mu/\sqrt{1-(\mu p_\mu)^2}\Big)\,.
\end{equation}
The operators $q$ and $p_\mu$ are self-adjoint and also symmetric with respect to the inner
product
\begin{eqnarray}\label{Snyd-IP}
\langle\psi_1|\psi_2\rangle_{\gamma_\mu,p_\mu}=\frac{\mu}{2}\int_{-1/\mu}^{+1/\mu}
\,\frac{dp_\mu\,\psi_1^{\ast}(p_{\mu})\psi_2(p_{\mu})}{\sqrt{1-(\mu p_\mu)^2}}
\end{eqnarray}
in which we have used the definition (\ref{PHS-IP}). The
completeness relation in this setup will be
\begin{eqnarray}\label{NIO}
1=\frac{\mu}{2}\int_{-1/\mu}^{+1/\mu}\,\frac{dp_\mu}{\sqrt{1-(\mu p_\mu)^2}}|p_\mu
\rangle\langle p_\mu|\,,
\end{eqnarray}
and the inner product of the momentum eigenvectors becomes
\begin{equation}\label{ME-IP}
\langle p_\mu|p'_\mu\rangle=\sqrt{1-(\mu p_\mu)^2}\delta(p_\mu-p'_\mu)\,.
\end{equation}
The eigenvalue problem for the position operator in the noncanonical
representation (\ref{Snyd-MHS}) takes the form of the following
differential equation
\begin{equation}\label{PEVP}
i\sqrt{1-(\mu p_{\mu})^2}\,\frac{\partial \psi(p_\mu)}{\partial p_\mu}=\lambda
\psi_\lambda(p_\mu),
\end{equation}
with solution
\begin{equation}\label{PEVP-S1}
\psi_\lambda(p_\mu)=c\,\exp\Big[-i\lambda\frac{\sin^{-1}(\mu p_\mu)}{\mu}\Big].
\end{equation}
The constant $c$ should be fixed by the normalization condition
\begin{eqnarray}\label{Normalization-c}
1=cc^{\ast}\frac{\mu}{2}\int_{-1/\mu}^{+1/\mu}\,\frac{dp_\mu}{\sqrt{1-(
\mu p_{\mu})^2}}=cc^{\ast}\frac{\pi}{2},
\end{eqnarray}
in which definition (\ref{Snyd-IP}) is used. Thus, the normalized
eigenvectors for the position operator in noncanonical polymeric
representation will be
\begin{equation}\label{PEVP-S}
\psi_\lambda(p_\mu)=\sqrt{\frac{2}{\pi}}\,\exp\Big[-i\lambda\frac{
\sin^{-1}(\mu p_\mu)}{\mu}\Big].
\end{equation}
One would be tempted to conclude that the plane waves of the
standard Schr\"{o}dinger representation
$\psi_\lambda(p)\sim\,e^{-i\lambda p}$, with $\lambda\in{\mathbb R}$
and $p \in(-\infty,+\infty)$ are recovered in the limit
$\mu\rightarrow\,0$ since the polymeric momentum $p_\mu$ reduces to
the standard unbounded canonical momentum $p$ in this limit.
However, since the approximation (\ref{Polymerization}) is reliable
for $\mu p\ll1$, one should be careful when taking the continuum
limit in polymer framework \cite{QPR2}. But, it can be shown that
the the continuum limit will be correctly recovered for a given
small value of the lattice spacing $\mu$ (e.g. $\mu\sim\,l_{_{\rm
Pl}}$) \cite{QPR22}. Also, the position eigenstates (\ref{PEVP-S})
should be orthogonal in momentum polarization. To see this, we
consider their inner product as
\begin{eqnarray}\label{SP-EVP}
\langle\psi_{\lambda'}|\psi_{\lambda}\rangle=\frac{\mu}{\pi}\int_{-1/\mu}^{+1/
\mu}\,\frac{\exp\Big[-i(\lambda-\lambda')\frac{\sin^{-1}(\mu p_\mu)}{\mu}\Big]
\,dp_\mu}{\sqrt{1-(\mu p_{\mu})^2}}=\frac{\sin\big(\pi(\lambda-
\lambda')/2\mu\big)}{\pi(\lambda-\lambda')/2\mu}=\delta_{\lambda,\lambda'}\,,
\hspace{1cm}
\end{eqnarray}
in agreement with our pervious general definition (\ref{PHS-IP}).
In the last equality we have used the fact that $\frac{\lambda-
\lambda'}{2\mu}=n\in{\mathbb Z}$ since $\lambda$ and $\lambda'$ are
the eigenvalues of the position operator which are restricted to
belong to the lattice $\gamma_\mu$ (\ref{lattice}). Thus, the
position eigenvectors leave the lattice (\ref{lattice}) invariant
and are naturally orthogonal in polymer framework. As we will
show, this is not the case for the theories based on the deformed
Heisenberg algebra and an extra assumption is needed to recover
the orthogonality of the position eigenvectors.

Unlike the standard Schr\"{o}dinger representation, the expectation
value of the kinetic term of the Hamiltonian operator in the polymer
framework converges. Indeed, with the help of (\ref{Snyd-IP}) we get
\begin{equation}\label{KT-C}
\langle\psi_\lambda|\frac{p_\mu^2}{2m}|\psi_\lambda\rangle=\frac{1}{4m\mu^2}\,.
\end{equation}
This result indicates the existence of an upper bound for the energy
of the free particle in polymer framework \cite{QPR2}. In contrast,
the expectation value of the kinetic term of the Hamiltonian
diverges in the minimal length uncertainty relation framework
\cite{GUP-HS}.

\subsection{The Modified Uncertainty Relation}

The position operator $q$ is a self-adjoint and symmetric operator
in polymer framework. Also, it is diagonal in the basis of the
eigenstates (\ref{PEVP-S}) which means that there is no uncertainty
in measurement of position. To be more precise, let us take a look
at the uncertainty principle in this setup.

The deformed Heisenberg algebra (\ref{CR-PL}) in the noncanonical
polymeric representation naturally provides a modification to the
standard uncertainty relation through the standard (quantization
scheme independent) definition
\begin{equation}\label{UR}
(\Delta A)^2\,(\Delta B)^2\geq\frac{1}{4}\langle[A,B]\rangle^2\,,
\end{equation}
for two arbitrary operators $A$ and $B$, where $\Delta
A\equiv\sqrt{\langle A^2\rangle-\langle A\rangle^2}$ and similarly
for $B$. For the noncanonical pair $(q,p_\mu)$ with commutation
relation (\ref{CR-PL}), the definition (\ref{UR}) gives
\begin{equation}\label{UR-PL}
\Delta q\,\Delta p_\mu\geq\frac{1}{2}\Big\langle\sqrt{1-(\mu
p_{\mu})^2}\Big\rangle,
\end{equation}
which is in agreement with the result obtained in Refs.
\cite{QPR,Hussain}. In the limit $\mu\rightarrow\,0$ we get
\begin{equation}\label{CR-PL-App}
\Delta q\,\Delta p_{\mu}\geq\frac{1}{2}\Big(1-\frac{\mu^2}{2}(\Delta p_{\mu})^2+
{\mathcal O}(\mu^4)\Big)\,,
\end{equation}
in which we have set $\langle p_{\mu}\rangle=0$. Although this
relation, in some senses, is very similar to the results obtained
from the generalized uncertainty relations which are investigated in
the context of the string theory \cite{GUP,GUP-HS,GUP-HS2}, there is
a crucial difference. The generalized uncertainty principle theories
predict a minimal (nonzero) uncertainty $\Delta q_0>0$ in position
measurement such that (see Ref. \cite{GUP-HS} for details)
\begin{equation}\label{MUIP}
(\Delta q)_{|\psi\rangle}=\langle\psi|(q-\langle\psi|q|\psi\rangle)^2
|\psi\rangle\geq\Delta q_0\,,
\end{equation}
which clearly implies that there cannot be any physical state which
is a position eigenvector since such an eigenstate should, of course,
have zero uncertainty in position . On the other hand, the
polymer-modified uncertainty relation (\ref{UR-PL}) does not predict
any minimal uncertainty in position measurement. Here, the existence
of a minimal length (the polymer length scale $\mu$) is encoded in
the discrete position eigenvalues rather than the minimal uncertainty
in position measurement. The advantage of this viewpoint to the
minimal length becomes more clear when we note that one can work
either in position or in momentum polarization in polymer framework.
But, the position polarization fails to be applicable in the
generalized uncertainty theories due to the existence of the nonzero
minimal uncertainty $\Delta q_0$ in position measurement \cite{GUP-HS}.

\subsection{Simple Harmonic Oscillator}

In this section we consider the well-known example of the harmonic
oscillator in the noncanonical polymeric representation of quantum
mechanic, whose energy eigenvalues are given by equation
(\ref{HEVP}). This eigenvalue problem may be solved in position or
momentum polarization and as we have mentioned before it becomes a
difference equation in position polarization while takes the form of
a differential equation in momentum polarization. In momentum
polarization, one can consider the eigenvalue problem (\ref{HEVP})
in canonical or noncanonical polymeric representations on the
Hilbert spaces (\ref{HS-S1}) and (\ref{Snyd-HS}), respectively. Let
us first consider this problem in the noncanonical representation
and then we show the equivalence of the results with those obtained
from the standard canonical representation.

In the noncanonical representation, the action of the position and
momentum operators on a typical wave function are defined by
relations (\ref{Snyd-MHS}) and it is straightforward to show that
the eigenvalue problem (\ref{HEVP}) for the harmonic oscillator with
potential $W=\frac{1}{2}m\omega^2q^2$ takes the form of the
following differential equation
\begin{equation}\label{Mathieu0}
\frac{d^2\psi(p_{\mu})}{dp_{\mu}^2}-\frac{\mu^2 p_{\mu}}{1-(\mu p_{\mu})^2}\frac{d\psi(
p_{\mu})}{dp_{\mu}}+\frac{d^4(\epsilon-p_{\mu}^2)}{1-(\mu p_{\mu})^2}\psi(p_{\mu})=0\,,
\end{equation}
where we have defined $d:=1/\sqrt{m\omega}$ and $\epsilon:=2mE$. Setting
\begin{eqnarray}\label{Mathieu1}
\phi:=\sin^{-1}(\mu p_\mu)+\frac{\pi}{2},
\end{eqnarray}
simplifies the above equation as
\begin{equation}\label{Mathieu}
\frac{d^2\psi(\phi)}{d\phi^2}+\big(a-2h^2\cos(2\phi)\big)\psi(\phi)=0\,,
\end{equation}
with
\begin{eqnarray}\label{Parameters}
a=2h(\epsilon{d^2}-h)\hspace{.4cm}\mbox{and}\hspace{.4cm}h=\frac{
d^2}{2\mu^2}.
\end{eqnarray}
The above differential equation is a Mathieu equation whose
solutions can be written in terms of the Mathieu cosin and sine
functions as \cite{Abramowitz}
\begin{eqnarray}\label{Mathieu2}
\psi(\phi)=c_1{\mathcal C}(a,h^2,\phi)+c_2{\mathcal S}(a,h^2,\phi),
\end{eqnarray}
where $c_1$ and $c_2$ are the constants of integration. Note that
the Fourier transform of the wave function (\ref{Mathieu2}) gives
the corresponding wave function in the position polarization which
clearly is periodic since it is restricted on the lattice
(\ref{lattice}) by definition. In this respect, the solution
(\ref{Mathieu2}) should be periodic to ensure appropriate fall-off
for the associated Fourier coefficient \cite{QPR} (see also
\cite{Abramowitz} for the properties of the periodic solution of the
Mathieu equation). In the limit $\mu/d\ll1$, one can use the
asymptotic formula $a_n=-2h^2+2h(2n+1)-\frac{
2n^2+2n+1}{4}+{\mathcal O}(h^{-1})$, to arrive at the following
energy eigenvalues
\begin{equation}\label{SHO-EV}
E_n=(2n+1)\frac{\omega}{2}-\frac{2n^2+2n+1}{4}\frac{\omega}{2}\,\Big(
\frac{\mu}{d}\Big)^2+{\mathcal O}\Big(\frac{\mu^4}{d^4}\Big).
\end{equation}
The first term in the right hand side of the above equation is
nothing but the energy eigenvalues of the harmonic oscillator in the
standard Schr\"{o}dinger representation. The second term is the
first polymeric correction although it is smaller than can be
detected \cite{QPR}. The energy eigenvalues (\ref{SHO-EV}) are in
agreement with the result obtained in Ref. \cite{QPR} from the
standard canonical polymeric representation of the quantum
mechanics. \footnote{The relation (\ref{SHO-EV}) exactly coincides
with the relation (5.10) of the Ref. \cite{QPR} just by substituting
$\mu=\mu_0/2$. This is because of the different definition of the
momentum squared (\ref{PLMO2}). More precisely, the squared of the
momentum (\ref{PLMO2}) shifts the states two steps in both sides in
the lattice (\ref{lattice}). But, the momentum squared that defined
in Ref. \cite{QPR} shifts the states one step in both sides.} This
coincidence refers to the fact that two Hilbert spaces (\ref{HS-S1})
and (\ref{Snyd-HS}) are unitarily equivalent. This is because they
are just the canonical and noncanonical representations of the
unique Hilbert space (\ref{HS-S1M}). Furthermore, the result
(\ref{SHO-EV}) is the same as the harmonic oscillator eigenvalues
which obtained in the Snyder-deformed Heisenberg algebra setup. This
coincidence signal a connection between these two apparently
different setup and we will explore this connection in the next
section.

\section{Connection with Noncommutative Geometry}

\subsection{Snyder-deformed Heisenberg Algebra}
In 1947, Snyder formulated a Lorentz invariant space-time which
admits a minimal length scale (ultraviolet cutoff) \cite{Snyder}.
Snyder's reasonable assumptions can be also translated into a
noncanonical Heisenberg algebra in order to achieve a minimal
length scale in standard quantum mechanics \cite{Mignemi}. In
this regard, we first introduce an Snyder-deformed Heisenberg
algebra ideas and then we will show how it can be identified
with the noncanonical polymeric representation of the quantum
mechanics. The strategy is based on the following arguments
\cite{Snyder-NPS}:
\begin{itemize}
\item Fixing the noncommutativity between coordinates as
\begin{eqnarray}\label{S-NCC}
[Q_i,Q_j]=\alpha^2\,J_{ij},
\end{eqnarray}
where $i,j=1,..,n$ and $\alpha$ is a deformation parameter with
dimension of length. This parameter is usually assumed to be of
order of the Planck length $l_{_{\rm Pl}}$ to ensure that these
effects becomes only important in the high energy regimes and also
being negligible in the limit of low energies (correspondence
principle). $J_{ij}=-J_{ji}=i(q_ip_j-q_jp_i)$ are the generators of
rotation in $n$ dimension. Such a form of the noncommutativity
between position coordinates is first proposed by Snyder himself
\cite{Snyder}.

\item The rotation generators should satisfy the ordinary $SO(n)$
algebra
\begin{eqnarray}\label{S-TG}
[J_{ij},J_{kl}]=\delta_{jk}J_{il}-\delta_{ik}J_{jl}-\delta_{jl}
J_{ik}-\delta_{il}J_{jk}\,.
\end{eqnarray}

\item The translation group is not deformed, i.e.,
\begin{eqnarray}\label{TG}
[P_i,P_j]=0.
\end{eqnarray}
This assumption is reasonable since we are looking just for the
ultraviolet effects such as minimal length and maximal momentum
(high energy regime modification). The noncommutativity between the
momentum operators is evidence of the existence the infrared cutoff
which is important only for the low energy regimes \cite{Mignemi}.
Therefore, since the ordinary momentum operators also satisfy the
commutation relations $[p_i,p_j]=0$, the momenta are the same either
in noncommutative or in commutative proposals.

\item In addition, it is natural to assume \cite{Snyder}
\begin{eqnarray}\label{S-RSP}
[J_{ij},Q_k]=Q_i\delta_{jk}-Q_j\delta_{ik}\\{[J_{ij},P_k]}=
P_i\delta_{jk}-P_j\delta_{ik}.
\end{eqnarray}
\end{itemize}
The above statements cannot fix the commutation relation between
variables $Q$ and $P$ and there are many noncommutative Heisenberg
algebras which satisfy all the above conditions and also are closed
in virtue of Jacobi identity. Nevertheless, relations (\ref{S-NCC})
and (\ref{S-RSP}) will be satisfied which are the only requirement
of the setup. It is always possible to find noncommutative
variables $(Q,P)$ in terms of the commutative pair $(q,p)$ such
that equation (\ref{S-NCC}) is satisfied (Darboux theorem). The
most general $SO(n)$ covariant form of such noncommutative
variables compatible with all of the above mentioned conditions is
\footnote{More precisely, the variables $(q,p)$ are Darboux
(canonical) variables and transformation (\ref{S-NCCD}) is a
Darboux transformation \cite{LT}.}
\begin{eqnarray}\label{S-NCCD}
Q_i=q_i\,{\varphi_1}+\alpha^2(q_jp_j)p_i\,\varphi_2\,,\nonumber\\
P_i=p_i\,,
\end{eqnarray}
where $\varphi_1=\varphi_1(\alpha^2{p^2})$ and
$\varphi_2=\varphi_2(\alpha^2 {p^2})$ are two unknown functions of
the momenta. The function $\varphi_1$ should satisfy the condition
$\varphi_1(0)=1$ for $\alpha\rightarrow\,0$ to recover the standard
commutation relations in the low energy regimes (correspondence
principle). Relations (\ref{S-NCCD}) give the commutation relation
between coordinates and momenta as
\begin{eqnarray}\label{S-NCCA}
[Q_i,P_j]=i(\delta_{ij}\varphi_1+\alpha^2 p_ip_j\varphi_2).
\end{eqnarray}
Our next challenge is how to determine two functions $\varphi_1$ and
$\varphi_2$. In fact, they should be obtained in such a way that the
relations (\ref{S-NCC}) and (\ref{S-RSP}) are satisfied.
Substituting (\ref{S-NCCD}) into the relation (\ref{S-RSP}), one can
see that this relation is automatically satisfied without any
restriction on the functions $\varphi_1$ and $\varphi_2$.
Substituting into the relation (\ref{S-NCC}), however, gives
\begin{eqnarray}\label{S-DE}
\frac{d\varphi_1}{d(\alpha^2{p^2})}+\frac{1}{2}\frac{1-\varphi_1\varphi_2
}{\varphi_1+\alpha^2{p^2}\varphi_2}=0.
\end{eqnarray}
It is easy to see that the commutation relations (\ref{S-NCCA}),
(\ref{S-NCC}), and (\ref{TG}) satisfy the Jacobi identity and
constitute the noncommutative Heisenberg algebra in this setup. This
noncommutative algebra can support a number of deformed Heisenberg
algebras just by choosing an appropriate form for the functions
$\varphi_1$ and $\varphi_2$ \cite{Snyder-NPS,Snyder-bounce}. Here,
we would like to construct a connection between the noncommutative
Heisenberg algebra (\ref{S-NCCA}) and the polymer quantum mechanic
and to do this, it is plausible to consider a one-dimensional system
for which the functions $\varphi_1$ and $\varphi_2$ are uniquely
fixed. Indeed, in this case we may find these functions as
$\varphi_1=\sqrt{1-{( \alpha p)}^2}$ and $\varphi_2=0$ leading to
the noncanonical Heisenberg algebra
\begin{eqnarray}\label{S-PNCA}
[Q,P]=i\sqrt{1-(\alpha P)^2},
\end{eqnarray}
in which the fact that $P=p$ is used. It is also important to note
that the sign of the deformation parameter $\alpha^2$ is not fixed
which may lead to different physical results. Evidently, the
momentum is bounded as $P\in(-\frac{1}{\alpha},+\frac{1}{\alpha})$
if $\alpha^2>0$ while a minimum nonzero uncertainty in position as
$\Delta Q_0= (\sqrt{-\alpha^2}/2)$ can be realized in the case of
$\alpha^2<0$ \cite{Snyder-bounce}. The former one is very similar to
the noncanonical polymeric representation with deformed Heisenberg
algebra (\ref{CR-PL}) and we only consider this case in this paper.

The Snyder-deformed Heisenberg algebra (\ref{S-PNCA}) together with the
associated Hamiltonian operator
\begin{equation}\label{S-Hamiltonian}
H=\frac{P^2}{2m}+W(Q),
\end{equation}
complete our construction to study the kinematics and dynamics of the
physical system in this setup. However, one should be careful that while the
Hamiltonian operator (\ref{S-Hamiltonian}) has the standard functional form,
the associated momentum is bounded as $P\in(-1/\alpha,+1/\alpha)$.

\subsection{Hilbert Space Representation}

Comparing the commutation relation (\ref{CR-PL}) with the
Snyder-deformed Heisenberg algebra (\ref{S-PNCA}), one immediately
recognizes a correspondence between momenta $p_\mu$ and $P$, they
satisfy the same modified Heisenberg algebra and also they are
bounded in the same way by the parameters $\mu$ and $\alpha$,
respectively. Thus, we identify the polymeric momentum $p_{\mu}$
with $P$ as
\begin{equation}\label{ID}
Q\equiv\,q,\hspace{.7cm}P\equiv\,p_\mu=\frac{1}{\mu}\sin(\mu p),
\end{equation}
in which we have also set the reasonable identification $\alpha=\mu$
for the deformation parameters. It is however important to note that
the polymer length $\mu$ is the lattice parameter and does not have
to always be very small although we consider it to be small compared
to other length scales present in the system under consideration. We
have done such assumption before in the study of harmonic oscillator
where we considered $\mu$ to be very small in comparison with the
natural length scale $d$. On the other hand, in the Snyder-deformed
Heisenberg algebra (\ref{S-PNCA}), the length scale $\alpha$ is
always assumed to be very small (for instance of order of the Planck
length) to induce a cutoff in ultraviolet regime
\cite{Snyder,Mignemi}. Thus, we suppose that the identification
(\ref{ID}) is possible for sufficiently small values of the polymer
length scale $\mu$. Here, we consider $\mu=\alpha={\mathcal O}(1)\,
l_{_{\rm Pl}}$ through the paper, where the numerical coefficient
${\mathcal O}(1)$ should be fixed by experiment
\cite{Gorji2,QGExperiment,PLMR-THR2}. Then, both of the polymer
quantum mechanics and the Snyder-deformed Heisenberg algebra
(\ref{S-PNCA}) can induce a maximal value for the momentum around
the Planck momentum $P_{_{\rm Pl}}=l_{_{\rm Pl}}^{-1}$ up to the
numerical factor ${\mathcal O}(1)$. There is also another point
here, that is, the polymer variables are restricted to belong to the
lattice (\ref{lattice}) while there is no such restriction when
dealing with the Snyder-deformed Heisenberg algebra (\ref{S-PNCA}).
However, as an extra structure, one may define a lattice also for
the Snyder-deformed case in order to get a self-adjoint position
operator with orthogonal eigenvectors \cite{Snyder-bounce}. In this
respect, as we will show, the position eigenvectors become naturally
orthogonal in the Snyder-deformed Heisenberg algebra when it
identifies with the noncanonical polymeric representation of the
quantum mechanics.

\begin{figure}
\flushleft\leftskip+12em{\includegraphics[width=3in]{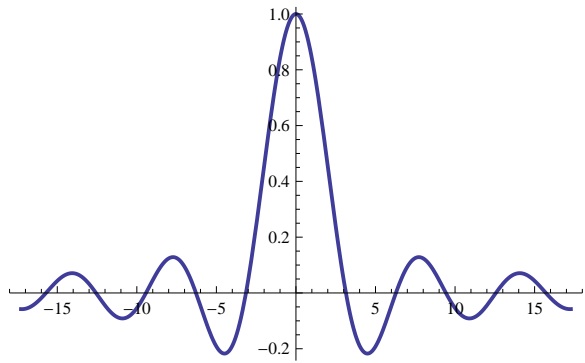}}
\hspace{3cm}\caption{\label{fig:1}
$\langle\psi_\lambda|\psi_{\lambda'} \rangle$ versus
$\lambda-\lambda'$ is plotted, where $\psi_\lambda$ denotes the
position eigenvector corresponds to the eigenvalue $\lambda$ in the
Snyder-deformed Heisenberg algebra (\ref{S-PNCA}) framework. The
figure shows that these eigenvectors are no longer orthogonal (see
relation (\ref{Snyd-EVP})). But, there is a family of orthogonal
eigenvectors for $\frac{\lambda-\lambda'}{2\alpha}\in{\mathbb Z}$.
Existence of this family of orthogonal eigenvectors indicates that
one should consider a lattice (such as (\ref{lattice}) which is
implemented in the polymer framework) to preserve the
orthogonality of the position eigenvectors. The figure is plotted
for $\alpha=\pi/2$.}
\end{figure}

As we say before, the eigenvalue problem of the harmonic oscillator
in the Snyder-deformed framework, gives the same result that we have
already obtained in the pervious section in the noncanonical
polymeric representation \cite{Snyder-bounce}. But, the inner
product of the position eigenstates (\ref{PEVP-S}) are no longer
generally orthogonal. Indeed, we have \cite{Snyder-bounce}
\begin{eqnarray}\label{Snyd-EVP}
\langle\psi_{\lambda'}|\psi_{\lambda}\rangle=\frac{\sin\big(\pi(
\lambda-\lambda')/2\alpha\big)}{\pi(\lambda-\lambda')/2\alpha},
\end{eqnarray}
and there is no reason to suppose $\frac{\lambda-\lambda'}{2\alpha
}\in{\mathbb Z}$ in this setup (see figure \ref{fig:1}). However,
one can define a family of orthogonal eigenvectors that make the
position operator diagonal \cite{GUP-HS,Snyder-bounce}. This is
equivalent to considering a lattice such as (\ref{lattice}) for
the position operator. Thus, the identification (\ref{ID}) is
reliable when one considers a lattice for the Snyder-deformed
Heisenberg algebra on a separable Hilbert space being isomorphic
to (\ref{HS-S1M}).

\section{Classical Phase Space}

Now, let us study the classical limit of the polymer quantum
mechanics which leads to a one-parameter family of $\mu$-dependent
classical theory rather than the ordinary classical one. The
standard classical theory would be recovered from the discrete
$\mu$-dependent one in the continuum limit $\mu\rightarrow\,0$.
However, one can also study such effective $\mu$-dependent theories
without any attribution to the quantum one by means of a process
known as {\it polymerization} in the literatures \cite{CPR}. The
method is based on implementing the Weyl operator on the classical
phase space in order to define an appropriate definition for the
discrete derivative of the phase space functions \cite{CPR,Gorji2}.
However, here, we should study the classical limit of the polymer
quantum mechanics which we have considered in the pervious sections.
More precisely, we analyze the classical phase space corresponds to
the Hilbert space (\ref{HS-S1M}) in two different pictures, i.e., in
canonical and noncanonical pictures correspond to the Hilbert spaces
(\ref{HS-S1}) and (\ref{Snyd-HS}), respectively. In this regard, the
momentum part of the phase space should be compactified as a circle
$S^1$ and the configuration space is the real line ${\mathbb R}$.
Thus, we consider a two-dimensional symplectic space $\Gamma_{\mu}$
with topology ${\mathbb R}\times S^1$ as the polymeric phase space
corresponds to the Hilbert space (\ref{HS-S1M}) and study its
classical implications for two canonical and noncanonical cases.

\subsection{Noncanonical Chart}

Let us start with the noncanonical representation of the polymeric
phase space $\Gamma_\mu$ and deal with the classical variables
$({\tilde q},{\tilde p_\mu})$ correspond to the operators
$(q,p_\mu)$, which are expected to satisfy the following Poisson
bracket
\begin{equation}\label{NC-PA}
\{{\tilde q},\,{\tilde p_{\mu}}\}=\sqrt{1-(\mu{\tilde p_{\mu}})^2}\,.
\end{equation}
The associated classical Hamiltonian function can easily be deduced
from the corresponding Hamiltonian operator (\ref{S-Hamiltonian}) as
\begin{equation}\label{NC-Hamiltonian}
H_{\mu}({\tilde q},{\tilde p_{\mu}})=\frac{{\tilde p_{\mu}}^2}{2m}
+W({\tilde q}).
\end{equation}
The Poisson algebra (\ref{NC-PA}) together with the Hamiltonian
function (\ref{NC-Hamiltonian}) determine the kinematics and
dynamics of the system on the polymeric phase space $\Gamma_\mu$.
Nevertheless, we implement the symplectic geometry in order to
define the associated Hamiltonian system which, as we will see,
gives a more clear connection between classical and quantum
frameworks for the polymeric systems.

Consider a symplectic manifold thought as the phase space
$\Gamma_\mu$ equipped with a symplectic structure $\omega$ which is
a closed nondegenerate $2$-form on $\Gamma_{\mu}$. We propose the
symplectic $2$-form being the noncanonical form
\begin{eqnarray}\label{NC-Structure}
\omega=\frac{d{\tilde q}\wedge\,d{\tilde p_{\mu}}}{
\sqrt{1-(\mu{\tilde p_{\mu}})^2}},
\end{eqnarray}
in this chart \cite{Gorji2}. The reason for this proposition will
become clear when we show that the symplectic structure
(\ref{NC-Structure}) correctly generates the noncanonical Poisson
algebra (\ref{NC-PA}). The Hamiltonian triplet is defined as
$(H_{\mu},\omega,{\bf x})$, where ${\bf x}$ is the Hamiltonian
vector field which determines the time evolution of the system
through the equation
\begin{equation}\label{VF-D}
i_{\bf x}\,\omega=dH_{\mu}\,.
\end{equation}
Substituting the symplectic structure (\ref{NC-Structure}) and the
associated Hamiltonian function (\ref{NC-Hamiltonian}) into the
equation (\ref{VF-D}) gives the solution for the vector field as
\begin{eqnarray}\label{NC-VF}
{\bf x}_{_H}=\sqrt{1-(\mu{\tilde p_{\mu}})^2}\Big(\frac{{\tilde
p_{\mu}}}{m}\frac{\partial}{\partial {\tilde q}}-\frac{\partial W}{
\partial{\tilde q}}\frac{\partial}{\partial{\tilde p_{\mu}}}\Big).
\end{eqnarray}
The integral curves of the vector field ${\bf x}$, in turn, give
the Hamilton's equation in this chart
\begin{eqnarray}\label{NC-HE}
\frac{d{\tilde q}}{dt}=\frac{\tilde p_{\mu}}{m}\,\sqrt{1-(\mu
{\tilde p_{\mu}})^2},\nonumber\\{\frac{d{\tilde p_{\mu}}}{dt}}
=-\frac{\partial W}{\partial{\tilde q}}\,\sqrt{1-(\mu
{\tilde p_{\mu}})^2},
\end{eqnarray}
which will be reduced to the standard Hamiltonian equations in the
limit $\mu\rightarrow \,0$. The Poisson bracket between two real
valued functions $F$ and $G$ on the phase space $\Gamma_{\mu}$ is
defined as
\begin{equation}\label{PB}
\{F,\,G\}=\omega({\bf x}_{_F},{\bf x}_{_G})\,,
\end{equation}
and closure of the symplectic structure ensures that the Jacobi
identity is satisfied by the resultant Poisson brackets. With the
help of the relations (\ref{NC-Structure}) and (\ref{NC-VF}), the
definition (\ref{PB}) gives the Poisson brackets in this chart as
\begin{eqnarray}\label{NC-PB}
\{F,\,G\}=\sqrt{1-(\mu{\tilde p_{\mu}})^2}\bigg(\frac{\partial F}{
\partial{\tilde q}}\frac{\partial G}{\partial{\tilde p_{\mu}}}-
\frac{\partial F}{\partial{\tilde p_{\mu}}}\frac{\partial G}{
\partial{\tilde q}}\bigg).
\end{eqnarray}
It is straightforward to check that the noncanonical Poisson algebra
can be retrieved from the relation (\ref{NC-PB}) by substituting
$F={\tilde q}$ and $G={\tilde p}$ which shows that the symplectic
structure (\ref{NC-Structure}) is chosen correctly.

\subsection{Darboux Chart}

According to the Darboux theorem, it is always possible to find a
local chart in which any symplectic structure takes the canonical
form. Thus, there is a (Darboux) transformation from the
noncanonical pair of variables $({\tilde q}, {\tilde p_{\mu}})$ to
a canonical one $({\tilde q},{\tilde p})$ in which the symplectic
$2$-form (\ref{NC-Structure}) takes the canonical form.

To see how this statement may work, consider the following Darboux
transformation on the phase space $\Gamma_{\mu}$
\begin{equation}\label{NC-Transformation}
({\tilde q},{\tilde p_{\mu}})\rightarrow({\tilde q},{\tilde p})=
\Big({\tilde q},\frac{1}{\mu}\sin^{-1}(\mu{\tilde p_{\mu}})\Big),
\end{equation}
which locally transforms the symplectic structure
(\ref{NC-Structure}) to the canonical form
\begin{eqnarray}\label{Dar-twoform}
\omega({\tilde q},{\tilde p_{\mu}})\rightarrow\omega({\tilde q},
{\tilde p})=d{\tilde q}\wedge\,d{\tilde p},
\end{eqnarray}
and also transforms the Hamiltonian function (\ref{NC-Hamiltonian})
to
\begin{equation}\label{class-hamiltonian}
H_{\mu}({\tilde q},{\tilde p_{\mu}})\rightarrow\,H_{\mu}({\tilde q}
,{\tilde p})=\,\frac{1}{m\mu^2}\Big(1\,-\,\cos(\mu{\tilde p})\Big)
+W({\tilde q}).
\end{equation}
This form for the Hamiltonian functional is the same as the
classical limit of the polymeric Hamiltonian operator
(\ref{qhamiltonian}). Substituting the canonical symplectic
structure (\ref{Dar-twoform}) and also the associated Hamiltonian
function (\ref{class-hamiltonian}) into the equation (\ref{VF-D}),
we are led to the following solution for the Hamiltonian vector
field
\begin{eqnarray}\label{Dar-VF}
{\bf x}_{_H}=\frac{\sin(\mu{\tilde p})}{m\mu}\frac{\partial}{
\partial{\tilde q}}-\frac{\partial W}{\partial{\tilde q}}\frac{
\partial}{\partial{\tilde p}}.
\end{eqnarray}
The integral curves of the above Hamiltonian vector field are the
polymer-modified Hamilton's equations of motion in the canonical
(Darboux) chart
\begin{equation}\label{Dar-PHE}
\frac{d{\tilde q}}{dt}=\frac{\sin(\mu{\tilde p})}{m\mu},\hspace{1cm}
\frac{d{\tilde p}}{dt}=-\frac{\partial W}{\partial {\tilde q}},
\end{equation}
which clearly reduce to the standard Hamilton's equations in the
continuum limit $\mu\rightarrow\,0$. It is straightforward to show
that the equations (\ref{Dar-PHE}) and (\ref{NC-HE}) are in
agreement through the Darboux
transformation (\ref{NC-Transformation}).

Substituting ${\bf x}_{_F}$ and ${\bf x}_{_G}$ from (\ref{Dar-VF})
into (\ref{PB}) one gets
\begin{eqnarray}\label{Dar-PB}
\{F,\,G\}=\frac{\partial F}{\partial{\tilde q}}\frac{
\partial G}{\partial{\tilde p}}-\frac{\partial F}{
\partial{\tilde p}}\frac{\partial G}{\partial{\tilde q}},
\end{eqnarray}
which is nothing but the standard definition for the canonical
Poisson bracket between two arbitrary functions $F$ and $G$.
Choosing $F({\tilde q},{\tilde p})= {\tilde q}$ and $G({\tilde
q},{\tilde p})={\tilde p}$, we can easily deduce
\begin{equation}\label{Dar-PA}
\{{\tilde q},\,{\tilde p}\}=1\,,
\end{equation}
which is coincided with the standard canonical Poisson algebra (\ref{CPA}).

Therefore, one deals with the unique Hamiltonian triplet
$(H_\mu,\omega, {\bf x})$ on the polymeric phase space $\Gamma_\mu$
which can be represented in two different local charts: i) In the
non-canonical chart, one can work with symplectic structure
(\ref{NC-Structure}), Hamiltonian function (\ref{NC-Hamiltonian}),
and the corresponding noncanonical Poisson algebra (\ref{NC-PA}).
ii) We can also work in the canonical chart and utilize the
effective Hamiltonian function (\ref{class-hamiltonian}), the
symplectic structure (\ref{Dar-twoform}) and the ordinary canonical
Poisson algebra (\ref{Dar-PA}). The former is the classical limit of
the noncanonical polymeric representation on the Hilbert space
(\ref{Snyd-HS}) and the latter is the classical limit of the
canonical representation on the Hilbert space (\ref{HS-S1}). The
trajectories on the polymeric phase space $\Gamma_\mu$ are the same
in either noncanonical and canonical charts since equation
(\ref{VF-D}) is hold in a chart-independent manner for the
Hamiltonian system $(H_\mu,\omega,{\bf x})$.

\section{Summary and Conclusions}

Existence of a minimal measurable length is a common feature of the
quantum gravity candidates such as loop quantum gravity and string
theory. This issue can also be realized from an algebraic
deformation in the standard quantum mechanical structures. The
so-called polymer quantum mechanics is investigated in the symmetric
sector of the loop quantum gravity which supports the existence of a
kind of minimal length scale $\mu$ known as the polymer length. On
the other hand, the generalized uncertainty relations have been
suggested in the context of string theory which also predict a
minimal length scale $\alpha$ (corresponds to the strings' size) for
the system under consideration. The minimal length scale is usually
taken to be of the order of Planck length and consequently either
the polymer quantum mechanics or the generalized uncertainty
theories induce an ultraviolet cutoff on Planck scale, where one
expects the quantum gravitational effects significantly would become
important. Therefore, we have considered the noncanonical Hilbert
space representation of the polymer quantum mechanics in order to
explore a possible connection between the polymer quantum mechanics
and the generalized uncertainty theories. All the physical result in
noncanonical representation, of course, are the same as those
obtained in the canonical representation since these representations
are unitarily equivalent. However, the relation with the generalized
uncertainty theories immediately emerges since the Heisenberg
algebra get modified in the noncanonical polymeric representation.
The polymer-modified Heisenberg algebra (\ref{CR-PL}) in the
noncanonical representation leads to the polymer-modified
uncertainty relation (\ref{UR-PL}). We have shown that the
polymer-modified Heisenberg algebra (\ref{CR-PL}) is coincided with
the Snyder-deformed (noncommutative) Heisenberg algebra
(\ref{S-PNCA}) in which the translation group is not deformed.
Technically, however, there is a crucial difference between the
polymer quantum mechanics and the generalized uncertainty theories.
In polymer framework, the eigenvalues of the position operator are
taken to be discrete with respect to the minimal (polymer) length
scale $\mu$. In the generalized uncertainty theories, however, the
minimal length scale is emerged through a nonzero minimal
uncertainty in position measurement which forbid the position
eigenvectors with zero uncertainty in position. The position
eigenvectors are no longer orthogonal and the position polarization
fails to be applicable in this setup. In contrast, the position
operator is restricted to belong to the lattice (\ref{lattice}) in
polymer framework and its eigenvectors are generally orthogonal and
leave the lattice invariant. Thus, both the position and momentum
polarization are relevant (self-adjoint) operators in polymer
framework. Therefore, the polymer viewpoint to the minimal length
scale is more fundamental. At classical level, the noncanonical
representation of the polymer quantum mechanics is determined by the
polymer-modified Poisson algebra (\ref{NC-PA}) and the Hamiltonian
function (\ref{NC-Hamiltonian}). In the canonical representation,
however, the Poisson algebra takes its standard canonical form
(\ref{Dar-PA}), but the corresponding Hamiltonian function get
modified as (\ref{class-hamiltonian}). We have shown that these two
pictures are determined by the unique Hamiltonian system and they
are related by a Darboux transformation on the corresponding phase
space. At the end, a natural question may arise: does it follow from
the comparison to Snyder space-time that polymer quantization could
be Lorentz invariant? In this sense we would like to emphasize that
since like the standard quantum mechanics, the polymer quantum
mechanics is not Lorentz invariant, it is just equivalent to the
quantum mechanics in Snyder noncommutative setup. However, the
Snyder noncommutative space-time may be equivalent with polymer
quantum field theory.

\end{document}